\documentclass[12pt]{article}
\usepackage[utf8x]{inputenc}
\usepackage{natbib}
\usepackage{fontenc}

\title{Unpacking the Social Media Bot: A Typology to Guide Research and Policy
}

\author{Robert Gorwa\footnote{Department of Politics and International Relations, University of Oxford. \texttt{@rgorwa}} \and Douglas Guilbeault\footnote{Annenberg School for Communication, University of Pennsylvania. \texttt{@dzguilbeault} 
\textit{\textit{Policy \& Internet, Fall 2018. This is a pre-publication version: please refer to final for page numbers/references. A draft of this paper was presented at ICA 2018, Prague (CZ).}} }}
\date{}

\begin{document}
\maketitle

\begin{abstract}
Amidst widespread reports of digital influence operations during major elections, policymakers, scholars, and journalists have become increasingly interested in the political impact of social media `bots.' Most recently, platform companies like Facebook and Twitter have been summoned to testify about bots as part of investigations into digitally-enabled foreign manipulation during the 2016 US Presidential election. Facing mounting pressure from both the public and from legislators, these companies have been instructed to crack down on apparently malicious bot accounts. But as this article demonstrates, since the earliest writings on bots in the 1990s, there has been substantial confusion as to exactly what a `bot' is and what exactly a bot does. We argue that multiple forms of ambiguity are responsible for much of the complexity underlying contemporary bot-related policy, and that before successful policy interventions can be formulated, a more comprehensive understanding of bots --- especially how they are defined and measured --- will be needed. In this article, we provide a history and typology of different types of bots, provide clear guidelines to better categorize political automation and unpack the impact that it can have on contemporary technology policy, and outline the main challenges and ambiguities that will face both researchers and legislators concerned with bots in the future. 
\end{abstract}

\section{Introduction}
\label{S:1}
The same technologies that once promised to enhance democracy are now
increasingly accused of undermining it. Social media services like
Facebook and Twitter---once presented as liberation technologies
predicated on global community and the open exchange of ideas---have
recently proven themselves especially susceptible to various forms of
political manipulation (Tucker et al. 2017). One of the leading
mechanisms of this manipulation is the social media ``bot,'' which has
become a nexus for some of the most pressing issues around algorithms,
automation, and Internet policy (Woolley and Howard 2016). In 2016
alone, researchers documented how social media bots were used in the
French elections to spread misinformation through the concerted
MacronLeaks campaign (Ferrara 2017), to push hyper-partisan news during
the Brexit referendum (Bastos and Mercea 2017), and to affect political
conversation in the lead up to the 2016 US Presidential election (Bessi
and Ferrara 2016). Recently, representatives from Facebook and Twitter
were summoned to testify before Congress as part of investigations into
digitally enabled foreign manipulation during the 2016 US Presidential
election, and leading international newspapers have extensively covered
the now-widely accepted threat posed by malicious bot accounts trying to
covertly influence political processes around the world. Since then, a
number of speculative solutions have been proposed for the so-called bot
problem, many of which appear to rely on tenuous technical capacities at
best, and others which threaten to significantly alter the rules
governing online speech, and at worst, embolden censorship on the behalf
of authoritarian and hybrid regimes. While the issues that we discuss in
this article are complex, it has become clear that the technology policy
decisions made by social media platforms as they pertain to automation,
as in other areas (Gillespie 2015), can have a resounding impact on
elections and politics at both the domestic and international level.

It is no surprise that various actors are therefore increasingly
interested in influencing bot policy, including governments,
corporations, and citizens. However, it appears that these stakeholders
often continue to talk past each other, largely due to a lack of basic
conceptual clarity. What exactly are bots? What do they do? Why do
different academic communities understand bots quite differently? The
goal of this article is to unpack some of these questions, and to
discuss the key challenges faced by researchers and legislators when it
comes to bot detection, research, and eventually, policy.

\subsection{An Overview of Ambiguities}

Reading about bots requires one to familiarize oneself with an
incredible breadth of terminology, often used seemingly interchangeably
by academics, journalists, and policymakers. These different terms
include: robots, bots, chatbots, spam bots, social bots, political bots,
botnets, sybils, and cyborgs, which are often used without precision to
refer to everything from automated social media accounts, to recommender
systems and web scrapers. Equally important to these discussions are
terms like trolling, sock-puppets, troll farms, and astroturfing
(Woolley 2016). According to some scholars, bots are responsible for
significant proportions of online activity, are used to game algorithms
and recommender systems (Yao et al. 2017), can stifle (Ferrara et al.
2016) or encourage (Savage, Monroy-Hernandez, and Hollerer 2015)
political speech, and can play an important role in the circulation of
hyperpartisan ``fake news'' (Shao et al. 2017). Bots have become a fact
of life, and to state that bots manipulate voters online is now accepted
as uncontroversial. But what exactly are bots?

Although it is now a commonly used term, the etymology of ``bot'' is
complicated and ambiguous. During the early days of personal computing,
the term was employed to refer to a variety of different software
systems, such as daemons and scripts that communicated warning messages
to human users (Leonard 1997). Other types of software, such as the
early programs that deployed procedural writing to converse with a human
user, were eventually referred to as ``bots'' or ``chatbots.'' In the
2000s, ``bot'' developed an entirely new series of associations in the
network and information security literatures, where it was used to refer
to computers compromised, co-opted, and remotely controlled by malware
(Yang et al. 2014). These devices can be linked in a network (a
``botnet'') and used to carry out distributed denial of service (DDoS)
attacks (Moore and Anderson 2012). Once Twitter emerged as a major
social network (and major home for automated accounts), some researchers
began calling these automated accounts ``bots,'' while others,
particularly computer scientists associated with the information
security community, preferred the term ``sybil''---a computer security
term that refers to compromised actors or nodes within a network (Alvisi
et al. 2013; Ferrara et al. 2016).

This cross-talk would not present such a pressing problem were not for
the policymakers and pundits currently calling for platform companies to
prevent foreign manipulation of social networks and to enact more
stringent bot policy (Glaser 2017). Researchers hoping to contribute to
these policy discussions have been hindered by a clear lack of
conceptual clarity, akin to the phenomenon known by social scientists as
concept misformation or category ambiguity (Sartori 1970). As Lazarsfeld
and Barton (1957) once argued, before we can investigate the presence or
absence of some concept, we need to know precisely what that concept is.
In other words, we need to better understand bots before we can really
research and write about them.

In this article, we begin by outlining a typology of bots, covering
early uses of the term in the pre-World Wide Web era up to the recent
increase in bot-related scholarship. Through this typology, we then go
on to demonstrate three major sources of ambiguity in defining bots: (1)
structure, which concerns the substance, design, and operation of the
``bot'' system, as well as whether these systems are algorithmically or
human-based; (2) function, which concerns how the ``bot'' system
operates over social media, for example, as a data scraper or an account
emulating a human user and communicating with other users; and (3) uses,
which concerns the various ways that people can use bots for personal,
corporate, and political ends, where questions of social impact are
front and center. We conclude with a discussion of the major challenges
in advancing a general understanding of political bots, moving forward.
These challenges include access to data, bot detection methods, and the
general lack of conceptual clarity that scholars, journalists, and the
public have had to grapple with.


\section{A Typology of Bots}
\label{S:2}

In its simplest form, the word ``bot'' is derived from ``robot.'' Bots
are have been generally defined as automated agents that function on an
online platform (Franklin and Graesser 1996). As some put it, these are
programs that run continuously, formulate decisions, act upon those
decisions without human intervention, and are able adapt to the context
they operate in (Tsvetkova et al. 2017). However, since the rise of
computing and the eventual creation of the World Wide Web, there have
been many different programs that have all been called bots, including
some that fulfill significantly different functions and have different
effects than those that we would normally associate with bots today. One
of the pioneering early works on bots, Leonard's \emph{Origin of New
Species} (1997), provides an excellent example of the lack of clarity
that the term had even as it first became widely used in the 1990s.
Various programs and scripts serving many different functions are all
lumped into Leonard's ``bot kingdom,'' such as web scrapers, crawlers,
indexers, interactive chatbots that interact with users via a simple
text interface, and the simple autonomous agents that played a role in
early online ``multi-user dungeon'' (MUD) games. Each one of these types
functions in different ways, and in recent years, has become associated
with a different scholarly community. While a complete typology would be
worthy of its own article, we provide here a brief overview of the major
different processes and programs often referred to as ``bots,'' paying
particular attention to those that are most relevant to current policy
concerns.

\subsection{`Web Robots': Crawlers and Scrapers}

As the Web grew rapidly following its inception in the 1990s, it became
clear that both accessing and archiving the incredible number of
webpages that were being added every day would be an extremely difficult
task. Given the unfeasibility of using manual archiving tools in the
long term, automated scripts---commonly referred to as robots or
spiders---were deployed to download and index websites in bulk, and
eventually became a key component of what are now known as search
engines (Olston and Najork 2010; Pant, Srinivasan, and Menczer 2004).

While these crawlers did not interact directly with humans, and operated
behind the scenes, they could still have a very real impact on
end-users: it quickly became apparent that these scripts posed a
technology policy issue, given that poorly executed crawlers could
inadvertently overwhelm servers by querying too many pages at once, and
because users and system administrators would not necessarily want all
of their content indexed by search engines. To remedy these issues, the
``Robot Exclusion Protocol'' was developed by the Internet Engineering
Task Force (IETF) to govern these ``Web Robots'' via a robots.txt file
embedded in webpages, which provided rules for crawlers as to what
should be considered off limits (Koster 1996). From their early days,
these crawlers were often referred to as bots: for example, Polybot and
IRLBot were two popular early examples (Olston and Najork 2010). Other
terminology used occasionally for these web crawlers included
``wanderers,'' ``worms,'' ``fish,'' ``walkers,'' or ``knowbots''
(Gudivada et al. 1997).

Today, it has become common to begin writing on social media bots with
big figures that demonstrate their apparent global impact. For example,
reports from private security and hosting companies have estimated that
more than half of all web traffic is created by ``bots,'' and these
numbers are occasionally cited by scholars in the field (Gilani,
Farahbakhsh, and Crowcroft 2017). But a closer look indicates that the
``bots'' in question are in fact these kinds of web crawlers and other
programs that perform crawling, indexing, and scraping functions. These
are an infrastructural element of search engines and other features of
the modern World Wide Web that do not directly interact with users on a
social platform, and are therefore considerably different than automated
social media accounts.

\subsection{Chatbots}
Chatbots are a form of human--computer dialog system that operate
through natural language via text or speech (Deryugina 2010; Sansonnet,
Leray, and Martin 2006). In other words, they are programs that
approximate human speech and interact with humans directly through some
sort of interface. Chatbots are almost as old as computers themselves:
Joseph Weizenbaum's program, ELIZA, which operated on an early
time-shared computing system at MIT in the 1960s, impersonated a
psychoanalyst by responding to simple text-based input from a list of
pre-programmed phrases (Weizenbaum 1966).

Developers of functional chatbots seek to design programs that can
sustain at least basic dialogue with a human user. This entails
processing inputs (through natural language processing, for example),
and making use of a corpus of data to formulate a response to this input
(Deryugina 2010). Modern chatbots are substantially more sophisticated
than their predecessors: today, chatbot programs have many commercial
implementations, and are often known as virtual assistants or assisting
conversational agents (Sansonnet, Leray, and Martin 2006), with current
voice-based examples including Apple's Siri and Amazon's Alexa. Another
implementation for chatbots is within messaging applications, and as
instant messaging platforms have become extremely popular, text-based
chatbots have been developed for multiple messaging apps, including
Facebook Messenger, Skype, Slack, WeChat, and Telegram (Folstad and
Brandtzaeg 2017). Bots have been built by developers to perform a range
of practical functions on these apps, including answering frequently
asked questions and performing organizational tasks. While some social
media bots, like those on Twitter, can occasionally feature chatbot
functionality that allows them to interact directly with human users
(see, for instance, the infamous case of Microsoft's ``Tay'' in Neff and
Nagy 2016), most chatbots remain functionally separate from typical
social media bots.

\subsection{Spambots}

Spam has been a long-standing frustration for users of networked
services, pre-dating the Internet on bulletin boards like USENET
(Brunton 2013). As the early academic ARPANET opened up to the general
public, commercial interests began to take advantage of the reach
provided by the new medium to send out advertisements. Spamming activity
escalated rapidly as the Web grew, to the point that spam was said to
``threaten the Internet's stability and reliability'' (Weinstein 2003).
As spam grew in scale, spammers wrote scripts to spread their messages
at scale---enter the first ``spambots.''

Spambots, as traditionally understood, are not simple scripts but rather
computers or other networked devices compromised by malware and
controlled by a third party (Brunton 2012). These have been
traditionally termed ``bots'' in the information security literature
(Moore and Anderson 2012). Machines can be harnessed into large networks
(botnets), which can be used to send spam en masse or perform
Distributed Denial of Service (DDoS) attacks. Major spam botnets, like
Storm, Grum, or Rostock, can send billions of emails a day and are
composed of hundreds of thousands of compromised computers
(Rodríguez-Gómez et al. 2013). These are machines commandeered for a
specific purpose, and not automated agents in the sense of a chatbot or
social bot (see below).

Two other forms of spam that users often encounter on the web and on
social networks are the ``spambots'' that post on online comment
sections, and those that spread advertisements or malware on social
media platforms. Hayati et al. (2009) study what they call ``web
spambots,'' programs that are often application specific and designed to
attack certain types of comment infrastructures, like the WordPress
blogging tools that provide the back-end for many sites, or comment
services like Disqus. These scripts function like a crawler, searching
for sites that accept comments and then mass posting messages. Similar
spam crawlers search the web to harvest emails for eventual spam emails
(Hayati et al. 2009). These spambots are effectively crawlers and are
distinct functionally from social bots. However, in a prime example of
the ambiguity that these terms can have, once social networking services
rose to prominence, spammers began to impersonate users with manually
controlled or automated accounts, creating profiles on social networks
and trying to spread commercial or malicious content onto sites like
MySpace (Lee, Eoff, and Caverlee 2011). These spambots are in fact
distinct from the commonly discussed spambots (networks of compromised
computers or web crawlers) and in some cases may only differ from
contemporary social media bots in terms of their use.

\subsection{Social Bots}
As the new generation of ``Web 2.0'' social networks were established in
the mid 2000s, bots became increasingly deployed on a host of new
platforms. On Wikipedia, editing bots were deployed to help with the
automated administration and editing of the rapidly growing crowdsourced
encyclopedia (Geiger 2014, 342). The emergence of the microblogging
service Twitter, founded in 2006, would lead to the large-scale
proliferation of automated accounts, due to its open application
programming interface (API) and policies that encouraged developers to
creatively deploy automation through third party applications and tools.
In the early 2010s, computer scientists began to note that these
policies enabled a large population of automated accounts that could be
used for malicious purposes, including spreading spam and malware links
(Chu et al. 2010).

Since then, various forms of automation operating on social media
platforms have been referred to as social bots. Two subtly different,
yet important distinctions have emerged in the relevant social and
computer science literatures, linked to two slightly different
spellings: ``socialbot'' (one word) and ``social bot'' (two words). The
first conference paper on socialbots'' published in 2011, describes how
automated accounts, assuming a fabricated identity, can infiltrate real
networks of users and spread malicious links or advertisements (Boshmaf
et al. 2011). These socialbots are defined in information security terms
as an adversary, and often called ``sybils,'' a term derived from the
network security literature for an actor that controls multiple false
nodes within a network (Cao et al. 2012; Boshmaf et al. 2013; Mitter,
Wagner, and Strohmaier 2014).

Social bots (two words) are a broader and more flexible concept,
generally deployed by the social scientists that have developed a recent
interest in various forms of automation on social media. A social bot is
generally understood as a program ``that automatically produces content
and interacts with humans on social media'' (Ferrara et al. 2016). As
Stieglitz et al. (2017) note in a comprehensive literature review of
social bots, this definition often includes a stipulation that social
bots mimic human users. For example, Abokhodair et al. (2015, 840)
define social bots as ``automated social agents'' that are public facing
and that seem to act in ways that are not dissimilar to how a real human
may act in an online space.

The major bot of interest of late is a subcategory of social bot: social
bots that are deployed for political purposes, also known as political
bots (Woolley and Howard 2016). One of the first political uses of
social bots was during the 2010 Massachusetts Special Election in the
United States, where a small network of automated accounts was used to
launch a Twitter smear campaign against one of the candidates (Metaxas
and Mustafaraj 2012). A more sophisticated effort was observed a year
later in Russia, where activists took to Twitter to mobilize and discuss
the Presidential election, only to be met with a concerted bot campaign
designed to clog up hashtags and drown out political discussion (Thomas,
Grier, and Paxson 2012). Since 2012, researchers have suggested that
social bots have been used on Twitter to interfere with political
mobilization in Syria (Abokhodair, Yoo, and McDonald 2015; Verkamp and
Gupta 2013) and Mexico (Suárez-Serrato et al. 2016), with journalistic
evidence of their use in multiple other countries (Woolley 2016). Most
recently, scholars have been concerned about the application of
political bots to important political events like referenda (Woolley and
Howard 2016), with studies suggesting that there may have been
substantial Twitter bot activity in the lead up to the UK's 2016 Brexit
referendum (Bastos and Mercea 2017), the 2017 French general election
(Ferrara 2017), and the 2016 US Presidential Election (Bessi and Ferrara
2016). While social bots are now often associated with state-run
disinformation campaigns, there are other automated accounts used to
fulfill creative and accountability functions, including via activism
(Savage, Monroy-Hernandez, and Hollerer 2015; Ford, Dubois, and
Puschmann 2016) and journalism (Lokot \&Diakopolous 2015). Social bots
can be used for both benign commercial purposes as well as more fraught
activities such as search engine optimization, spamming, and influencer
marketing (Ratkiewicz et al. 2011).

\subsection{Sockpuppets and `Trolls'}
The term ``sockpuppet'' is another term that is often used to describe
fake identities used to interact with ordinary users on social networks
(Bu, Xia, and Wang 2013). The term generally implies manual control over
accounts, but it is often used to include automated bot accounts as well
(Bastos and Mercea 2017). Sockpuppets can be deployed by government
employees, regular users trying to influence discussions, or by
``crowdturfers,'' workers on gig-economy platforms like Fiverr hired to
fabricate reviews and post fake comments about products (Lee, Webb, and
Ge 2014).

Politically motivated sockpuppets, especially when coordinated by
government proxies or interrelated actors, are often called ``trolls.''
Multiple reports have emerged detailing the activities of a legendary
troll factory linked to the Russian government and located outside of St
Petersburg, allegedly housing hundreds of paid bloggers who inundate
social networks with pro-Russia content published under fabricated
profiles (Chen 2015). This company, the so-called ``Internet Research
Agency,'' has further increased its infamy due to Facebook and Twitter's
recent congressional testimony that the company purchased advertising
targeted at American voters during the 2016 Presidential election
(Stretch 2017). There are varying degrees of evidence for similar
activity, confined mostly to the domestic context and carried out by
government employees or proxies, with examples including countries like
China, Turkey, Syria, and Ecuador (King et al. 2017; Cardullo 2015;
Al-Rawi 2014; Freedom House 2016).

The concept of the ``troll farm'' is imprecise due to its differences
from the practice of ``trolling'' as outlined by Internet scholars like
Phillips (2015) and Coleman (2012). Also challenging are the differing
cultural contexts and understandings of some of these terms.
Country-specific work on digital politics has suggested that the lexicon
for these terms can vary in different countries: for instance, in
Polish, the terms ``troll'' and ``bot'' are generally seen by some as
interchangeable, and used to indicate manipulation without regard to
automation (Gorwa 2017). In the public discourse in the United States
and United Kingdom around the 2016 US Election and about the Internet
Research Agency, journalists and commentators tend to refer to Russian
trolls and Russian bots interchangeably. Some have tried to get around
these ambiguous terms: Bastos and Mercea (2017) use the term sockpuppet
instead, noting that most automated accounts are in a sense
sockpuppets, as they often impersonate users. But given that the notion
of simulating the general behavior of a human user is inherent in the
common definition of social bots (Maus 2017), we suggest that automated
social media accounts be called social bots, and that the term
sockpuppet be used (instead of the term troll) for accounts with manual
curation and control.

\subsection{Cyborgs and Hybrid Accounts}

Amongst the most pressing challenges for researchers today are accounts
which exhibit a combination of automation and of human curation, often
called ``cyborgs.'' Chu et al. (2010, 21) provided one of the first, and
most commonly implemented definitions of the social media cyborg as a
``bot-assisted human or human-assisted bot.'' However, it has never been
clear exactly how much automation makes a human user a cyborg, or how
much human intervention is needed to make a bot a cyborg, and indeed,
cyborgs are very poorly understood in general. Is a user that makes use
of the service Tweetdeck (which was acquired by Twitter in 2011, and is
widely used) to schedule tweets or to tweet from multiple accounts
simultaneously considered a cyborg? Should organizational accounts (from
media organizations like the BBC, for example) which tweet automatically
with occasional human oversight be considered bots or cyborgs?

Another ambiguity regarding hybrids is apparent in the emerging trend of
users volunteering their real profiles to be automated for political
purposes, as seen in the 2017 UK general election (Gorwa and Guilbeault
2017). Similarly, research has documented the prevalence of underpaid,
human ``clickworkers'' hired to spread political messages and to like,
upvote, and share content algorithms (Lee et al., 2011, 2014).
Clickworkers offer a serviceable alternative to automated processes,
while also exhibiting enough human-like behavior to avoid anti-spam
filters and bot detection algorithms (Golumbia 2013). The conceptual
distinction between social bots, cyborgs, and sock-puppets is unclear,
as it depends on a theoretical and hereto undetermined threshold of
automation. This lack of clarity has a real effect: problematically, the
best current academic methods for Twitter bot detection are not able to
accurately detect cyborg accounts, as any level of human engagement is
enough to throw off machine-learning based models based on account
features (Ferrara et al. 2016). 

\section{A Framework for Understanding Bots: Three Considerations}
\label{S:3}

The preceding sections have outlined the multitude of different bots,
and the challenges of trying to formulate static definitions. When
creating a conceptual map or typology, should we lump together types of
automation by their use, or by how they work? Rather than attempting to
create a definitive, prescriptive framework for the countless different
types of bots, we recommend three core considerations that are useful
when thinking about them, inspired by past work on developer--platform
relations and APIs (Bogost \& Montfort 2008). Importantly, these
considerations are not framed as a rejection of pre-existing
categorizations, and they account for the fact that bots are constantly
changing and increasing in their sophistication. The framework has three
parts, which can be framed as simple questions. The idea is that
focusing on each consideration when assessing a type of bot will provide
a more comprehensive sense of how to categorize the account, relative to
one's goals and purposes. The first question is structural: How does the
technology actually work? The second is functional: What kind of
operational capacities does the technology afford? The third is ethical:
How are these technologies actually deployed, and what social impact do
they have? We discuss these three considerations, and their implications
for policy and research, below.

\subsection{The Structure of the System}
The first category concerns the substance, design, and operation of the
system. There are many questions that need to be considered. What
environment does it operate in? Does it operate on a social media
platform? Which platform or platforms? How does the bot work? What type
of code does it use? Is it a distinct script written by a programmer, or
a publicly available tool for automation like If This Then That (IFTTT),
or perhaps a type of content management software like SocialFlow or
Buffer? Does it use the API, or does it use software designed to
automate web-browsing by interacting with website html and simulating
clicks (headless browsing)? Is it fully automated, or is it a hybrid
account that keeps a ``human in the loop''? What type of algorithm does
it use? Is it strictly procedural (e.g. has a set number of responses,
like ELIZA) or does it use machine learning to adapt to conversations
and exhibit context sensitivity (Adams 2017)? Policy at both the
industry and public level will need to be designed differently to target
``bots'' with different structural characteristics.

Perhaps the simplest and most important question about structure for bot
regulation is whether the ``bot'' is made of software at all, or if it
is a human exhibiting bot-like behavior. A surprising number of
journalists and researchers describe human-controlled accounts as bots:
for example, Munger's (2017) online experiment where the so-called bot
accounts were manually controlled by the experimenter. Similarly, the
recent media coverage of ``Russian bots'' often lumps together automated
accounts and manually controlled ones under a single umbrella (Shane
2017). Even more ambiguous are hybrid accounts, where users can easily
automate their activity using various types of publicly available
software. At the structural level, technology policy will have to
determine how this type of automation will be managed, and how these
types of content management systems should be designed. The structure of
the bot is also essential for targeting technical interventions, either
in terms of automated detection and removal, or in terms of prevention
via API policies. If policy makers are particularly concerned with bots
that rely on API access to control and operate accounts, then lobbying
social media companies to impose tighter constraints on their API could
be an effective redress. Indeed, it appears as if most of the Twitter
bots that can be purchased online or through digital marketing agencies
are built to rely on the public API, so policy interventions at this
level are likely to lead to a significant reduction in bot activity.
Similarly, structural interventions would include a reshaping of how
content management allows the use of multiple accounts to send duplicate
messages and schedule groups of posts ahead of time.

\subsection{The Bot's Function}
The second category pertains more specifically to what the bot does. Is
the role of the bot to operate a social media account? Does it identify
itself as a bot, or does it impersonate a human user, and if so, does it
do so convincingly? Does it engage with users in conversation? Does it
communicate with individual users, or does it engage in unidirectional
mass-messaging?

Questions concerning function are essential for targeting policy to
specific kinds of bots. They are also vital for avoiding much of the
cross-talk that occurs in bot-related discourse. For instance, chatbots
are occasionally confused with other types of social bots, even though
both exhibit distinct functionalities, with different structural
underpinnings. In their narrow, controlled environment, chatbots are
often clearly identified as bots, and they can perform a range of
commercial services such as making restaurant reservations or booking
flights. Some chatbots have even been designed to build personal
relationships with users---such as artificial companies and therapist
bots (Floridi 2014; Folstad and Brandtzaeg 2017).

These new self-proclaimed bots pose their own issues and policy
concerns, such as the collection and marketing of sensitive personal
data to advertisers (Neff and Nafus 2016). Importantly, chatbots differ
substantially in both structure and function from most social bots,
which communicate primarily over public posts that appear on social
media pages. These latter bots are typically built to rely on hard-coded
scripts that post predetermined messages, or that copy the messages of
users in a predictable manner, such that they are incapable of
participating in conversations. Questions about functionality allow us
to distinguish social bots, generally construed, from other algorithms
that may not fall under prospective bot-related policy interventions
aimed at curbing political disinformation. If the capacity to
communicate with users is definitive of the type of bot in question,
where issues of deception and manipulation are key, then algorithms that
do not have direct public interaction with users should not be
considered to be conceptually similar; for example, web-scrapers,

\subsection{The Bot's Use}

This third category specifically refers to how the bot is used, and what
the end goal of the bot is. This is arguably the most important from a
policy standpoint, as it contains ethical and normative judgements as to
what positive, acceptable online behavior is---not just for bots, but
also for users in general. Is the bot being used to fulfil a political
or ideological purpose? Is it spreading a certain message or belief? If
so, is its goal designed to empower certain communities or promote
accountability and transparency? Or instead, does the bot appear to have
a commercial agenda?

Because of the diversity of accounts that qualify as bots, automation
policies cannot operate without normative assumptions about what kinds
of bots should be allowed to operate over social media. The problem for
the policymakers currently trying to make bots illegal (see, for
  example, the proposed ``Bot Disclosure and Accountability Act, 2018,'' also known as the Feinstein Bill). is that structurally, the same social bots can simultaneously enable a host of positive and negative actors. The affordances that make social bots a potentially powerful political organizing tool are the same ones
that allow for their implementation by foreign governments (for example), much like social networks themselves, and other recent digital technologies with similar ``dual-use'' implications (Pearce 2015). Therefore, it is difficult to constrain negative uses without also curbing positive uses at the structural level.

For instance, if social media platforms were to ban bots of all kinds as
a way of intervening on political social bots, this could prevent the
use of various chat bot applications that users appreciate, such as
automated personal assistants and customer service bots. Otherwise, any
regulation on bots, either from within or outside of social media
companies, would need to distinguish types of bots based on their
function in order to formulate clear regulations to address the types of
bots that have negative impact, while preserving the bots that are
recognized as having a more positive impact. As specified by the
topology above, it may be most useful to develop regulations to address
social bots particularly, given that webscrapers are not designed to
influence users through direct communicative activities, and chatbots
are often provided by software companies to perform useful social
functions.

The issue of distinguishing positive from negative uses of bots is
especially complex when considering that social media companies often
market themselves as platforms that foster free speech and political
conversation. If organizations and celebrities are permitted certain
types of automation---including those who use it to spread political
content---then it seems fair that users should also be allowed to deploy
bots that spread their own political beliefs. Savage et al. (2015), for
instance, have designed a system of bots to help activists in Latin
America mobilize against corruption. As political activity is a core
part of social media, and some accounts are permitted automation, the
creators of technology policy (most critically, the employees of social
media platforms who work on policy matters) will be placed in the
difficult position of outlining guidelines that do not arbitrarily
disrupt legitimate cases, such as citizen-built bot systems, in their
attempt to block illegitimate political bot activity, such as
manipulative foreign influence operations. But it is clear that
automation policies---like other content policies---should be made more
transparent, or they will appear wholly arbitrary or even purposefully
negligent. A recent example is provided by the widely covered account of
ImpostorBuster, a Twitter bot built to combat antisemitism and hate
speech, which was removed by Twitter, rather than the hate-speech bots
and accounts it was trying to combat (Rosenberg 2017). While Twitter is
not transparent as to why it removes certain accounts, it appears to
have been automatically pulled down for structural reasons (such as
violating the rate-limit set by Twitter, after having been flagged by
users trying to take the bot down) without consideration of its
normative use and possible social benefit.

Overall, it is increasingly evident that the communities empowered by
tools such as automation are not always the ones that the social media
platforms may have initially envisioned when they hoped that users would
use the tools---with the sophisticated use of bots, sock-puppets, and
other mechanisms for social media manipulation by the US ``alt-right''
in the past two years providing an excellent example (Marwick and Lewis
2017). Should social media companies crack down on automated accounts?
As platforms currently moderate what they consider to be acceptable
bots, a range of possible abuses of power become apparent as soon as
debates around disinformation and ``fake news'' become politicized. Now
that government interests have entered the picture, the situation has
become even more complex. Regimes around the world have already begun to
label dissidents as ``bots'' or ``trolls,'' and dissenting speech as
``fake news''---consider the recent efforts by the government of Vietnam
to pressure Facebook to remove ``false accounts'' that have espoused
anti-government views (Global Voices 2017). It is essential that social
media companies become more transparent about how they define and
enforce their content policies---and that they avoid defining bots in
such a vague way that they can essentially remove any user account
suspected of demonstrating politically undesirable behavior.

\section{Current Challenges for Bot-Related Policy}
\label{S:4}

Despite mounting concern about digital influence operations over social
media, especially from foreign sources, there have yet to be any
governmental policy interventions developed to more closely manage the
political uses of social media bots. Facebook and Twitter have been
called to testify to Congressional Intelligence Committees about bots
and foreign influence during the 2016 US presidential election, and have
been pressed to discuss proposed solutions for addressing the issue.
Most recently, measures proposed by state legislators in California in
April 2018, and at the federal level by Senator Diane Feinstein in June
2018, would require all bot accounts to be labeled as such by social media
companies (Wang 2018). However, any initiatives suggested by
policymakers and informed by research will have to deal with several
pressing challenges: the conceptual ambiguity outlined in the preceding
sections, as well as poor measurement and data access, lack of clarity
about who exactly is responsible, and the overarching challenge of
business incentives that are not predisposed towards resolving the
aforementioned issues.

\subsection{Measurement and Data Access}

Bot detection is very difficult. It is not a widely reported fact that
researchers are unable to fully represent the scale of the current issue
by relying solely on data provided through public APIs. Even the social
media companies themselves find bot detection a challenge, partially
because of the massive scale on which they (and the bot operators)
function. In a policy statement following its testimony to the Senate
Intelligence Committee in November 2017, Twitter said it had suspended
over 117,000 ``malicious applications'' in the previous four months
alone, and was catching more than 450,000 suspicious logins per day
(Twitter Policy 2017). Tracking the thousands of bot accounts created
every day, when maintaining a totally open API, is virtually impossible.
Similarly, Facebook has admitted that their platform is so large (with
more than two billion users) that accurately classifying and measuring
``inauthentic'' accounts is a major challenge (Weedon, Nuland, and
Stamos 2017). Taking this a step further by trying to link malicious
activity to a specific actor (e.g. groups linked to a foreign
government) is even more difficult, as IP addresses and other indicators
can be easily spoofed by determined, careful operators.

For academics, who do not have access to more sensitive account
information (such as IP addresses, sign-in emails, browser
fingerprints), bot detection is even more difficult. Researchers cannot
study bots on Facebook, due to the limitations of the publicly available
API, and as a result, virtually all studies of bot activity have taken
place on Twitter (with the notable exception of studies where
researchers have themselves deployed bots that invade Facebook, posing a
further set of ethical dilemmas, see Boshmaf et al. 2011). Many of the
core ambiguities in bot detection stem from what can be termed the
``ground truth'' problem: even the most advanced current bot detection
methods hinge on the successful identification of bot accounts by human
coders (Subrahmanian et al. 2016), a problem given that humans are not
particularly good at identifying bot accounts (Edwards et al. 2014).
Researchers can never be 100 percent certain that an account is truly a
bot, posing a challenge for machine learning models that use
human-labeled training data (Davis et al. 2016). The precision and
recall of academic bot detection methods, while constantly improving, is
still seriously limited. Less is known about the detection methods
deployed by the private sector and contracted by government agencies,
but one can assume that they suffer from the same issues.

Just like researchers, governments have data access challenges. For
example, what really was the scale of bot activity during the most
recent elections in the United States, France, and Germany? The key
information about media manipulation and possible challenges to
electoral integrity is now squarely in the private domain, presenting
difficulties for a public trying to understand the scope of a problem
while being provided with only the most cursory information. The policy
implications of these measurement challenges become very apparent in the
context of the recent debate over a host of apparently Russian-linked
pages spreading inflammatory political content during the 2016 US
presidential election. While Facebook initially claimed that only a few
million people saw advertisements that had been generated by these
pages, researchers used Facebook's own advertising tools to track the
reach that these posts had generated, concluding that they had been seen
more than a hundred million times (Albright 2017). However, Karpf (2017)
and others suggested that these views could have been created by
illegitimate automated accounts, and that there was no way of telling
how many of the ``impressions'' were from actual Americans. It is
currently impossible for researchers to either discount or confirm the
extent that indicators such as likes and shares are being artificially
inflated by false accounts, especially on a closed platform like
Facebook. The existing research that has been conducted by academics
into Twitter, while imperfect, has at least sought to understand what is
becoming increasingly perceived as a serious public interest issue.
However, Twitter has dismissed this work by stating that their API does
not actually reflect what users see on the platform (in effect, playing
the black box card). This argument takes the current problem of
measurement a step further: detection methods which are already
imperfect operate on the assumption that the Twitter Streaming APIs
provide a fair account of content on the platform. To understand the
scope and scale of the problem, policymakers will need more reliable
indicators and better measurements than are currently available.

\subsection{Responsibility}
Most bot policy to date has in effect been entirely the purview of
social media companies, who understandably are the primary actors in
dealing with content on their platforms and manage automation based on
their own policies. However, the events of the past year have
demonstrated that these private (often rather opaque) policies can have
serious political ramifications, potentially placing them more squarely
within the remit of regulatory and legal authorities. A key, and
unresolved challenge for policy is the question of responsibility, and
the inter-related questions of jurisdiction and authority. To what
extent should social media companies be held responsible for the
dealings of social bots? And who will hold these companies to
account?

While the public debate around automated accounts is only nascent at
best, it is clearly related to the current debates around the governance
of political content and hyper-partisan ``fake news.'' In Germany, for
instance, there has been substantial discussion around newly enacted
hate-speech laws which impose significant fines against social media
companies if they do not respond quickly enough to illegal content,
terrorist material, or harassment (Tworek 2017). Through such measures,
certain governments are keen to assert that they do have jurisdictional
authority over the content to which their citizens are exposed. A whole
spectrum of regulatory options under this umbrella exist, with some
being particularly troubling. For example, some have argued that the
answer to the ``bot problem'' is as simple as implementing and enforcing
strict ``real-name'' policies on Twitter---and making these policies
stricter for Facebook (Manjoo and Roose 2017). The recent emergence of
bots into the public discourse has re-opened age old debates about
anonymity and privacy online (boyd 2012; Hogan 2012), now with the added
challenge of balancing the anonymity that can be abused by sock-puppets
and automated fake accounts, and the anonymity that empowers activists
and promotes free speech around the world.

In a sense, technology companies have already admitted at least some
degree of responsibility for the current political impact of the
misinformation ecosystem, within which bots play an important role (Shao
et al. 2017). In a statement issued after Facebook published evidence of
Russian-linked groups that had purchased political advertising through
Facebook's marketing tools, CEO Mark Zuckerberg mentioned that Facebook
takes political activity seriously and was ``working to ensure the
integrity of the {[}then upcoming{]} German elections'' (Read 2017).
This kind of statement represents a significant acknowledgement of the
political importance of social media platforms, despite their past
insistence that they are neutral conduits of information rather than
media companies or publishers (Napoli and Caplan 2017). It is entirely
possible that Twitter's policies on automation have an effect, no matter
how minute, on elections around the world. Could they be held liable for
these effects? At the time of writing, the case has been legislated in
the court of public opinion, rather than through explicit policy
interventions or regulation, but policymakers (especially in Europe)
have continued to put Twitter under serious pressure to provide an
honest account of the extent that various elections and referenda (e.g.
Brexit) have been influenced by ``bots.'' The matter is by no means
settled, and will play an important part in the deeper public and
scholarly conversation around key issues of platform responsibility,
governance, and accountability (Gillespie 2018).

\subsection{Contrasting Incentives}

Underlying these challenges is a more fundamental question about the
business models and incentives of social media companies. As Twitter has
long encouraged automation by providing an open API with very permissive
third-party application policies, automation drives a significant amount
of traffic on their platform (Chu et al. 2010). Twitter allows accounts
to easily deploy their own applications or use tools that automate their
activity, which can be useful: accounts run by media organizations, for
example, can automatically tweet every time a new article is published.
Automated accounts appear to drive a significant portion of Twitter
traffic (Gilani et al. 2017; Wojcik et al. 2018), and indeed, fulfill
many creative, productive functions alongside their malicious ones.
Unsurprisingly, Twitter naturally wishes to maintain the largest
possible user base, and reports ``monthly active users'' to its
shareholders, and as such, is loath to change its automation policies
and require meaningful review for applications. It has taken immense
public pressure for Twitter to finally start managing the developers who
are allowed to build on the Twitter API, announcing a new ``developer
onboarding process'' in January 2018 (Twitter Policy 2018).

As business incentives are critical in shaping content policy---and
therefore policies concerning automation---for social media companies,
slightly different incentives have yielded differing policies on
automation and content. For example, while Twitter's core concern has
been to increase their traffic and to maintain as open a platform as
possible (famously once claiming to be the ``free speech wing of the
free speech party''), Facebook has been battling invasive spam for years
and has much tighter controls over its API. As such, it appears that
Facebook has comparatively much lower numbers of automated users (both
proportionally and absolutely), but, instead, is concerned primarily
with manually controlled sock-puppet accounts, which can be set up by
anyone and are difficult or impossible to detect if they do not
coordinate at scale or draw too much attention (Weedon, Nuland, and
Stamos 2017). For both companies, delineating between legitimate and
illegitimate activity is a key challenge. Twitter would certainly prefer
to be able to keep their legitimate and benign forms of automation (bots
which regularly tweet the weather, for example) and only clamp down on
malicious automation, but doing so is difficult, as the same structural
features enable both types of activity. These incentives seem to inform
the platforms' unwillingness to share data with the public or with
researchers, as well as their past lack of transparency. Evidence that
demonstrated unequivocally the true number of automated accounts on
Twitter, for example, could have major, adverse effects on their bottom
line. Similarly, Facebook faced public backlash after a series of
partnerships with academics that yielded unethical experiments
(Grimmelmann 2015). Why face another public relations crisis if they can
avoid it?

This illustrates the challenge that lies behind all the other issues we
have mentioned here: platform interests often clash with the preferences
of the academic research community and of the public. Academics strive
to open the black box and better understand the role that bots play in
public debate and information diffusion, while pushing for greater
transparency and more access to the relevant data, with little concern
for the business dealings of a social networking platform. Public
commentators may wish for platforms to take a more active stance against
automated or manually orchestrated campaigns of hate speech and
harassment, and may be concerned by the democratic implications of
certain malicious actors invisibly using social media, without
necessarily worrying about how exactly platforms could prevent such
activity, or the implications of major interventions (e.g. invasive
identity-verification measures). There are no easy solutions to these
challenges, given the complex trade-offs and differing stakeholder
incentives at play.

While scholars strive to unpack the architectures of contemporary media
manipulation, and legislators seek to understand the impact of social
media on elections and political processes, the corporate actors
involved will naturally weigh disclosures against their bottom line and
reputations. For this reason, the contemporary debates about information
quality, disinformation, and ``fake news''---within which lie the
questions of automation and content policy discussed in this
article---cannot exist separately from the broader debates about
technology policy and governance. Of the policy and research challenges
discussed in this last section, this is the most difficult issue moving
forward: conceptual ambiguity can be reduced by diligent scholarship,
and researchers can work to improve detection models, but business
incentives will not shift on their own. As a highly political, topical,
and important technology policy issue, the question of political
automation raises a number fundamental questions about platform
responsibility and governance that have yet to be fully explored by
scholars.

\section{Conclusion}
\label{S:5}

Amidst immense public pressure, policymakers are trying to understand
how to respond to the apparent manipulation of the emerging
architectures of digitally enabled political influence. Admittedly, the
debate around bots and other forms of political automation is only in
its embryonic stages; however, we predict that it will be a far more
central component of future debates around the political implications of
social media, political polarization, and the effects of ``fake news,''
hoaxes, and misinformation. For this to happen, however, far more work
will be needed to unpack the conceptual mishmash of the current bot
landscape. A brief review of the relevant scholarship shows that the
notion of what exactly a ``bot'' is remains vague and ill-defined. Given
the obvious technology policy challenges that these ambiguities present,
we hope that others will expand on the basic framework presented here
and continue the work through definitions, typologies, and conceptual
mapping exercises.

Quantitative studies have recently made notable progress in the ability
to identify and measure bot influence on the diffusion of political
messages, providing promising directions for future work (Vosoughi et
al. 2018). However, we expect that to maximize the benefits of these
studies for developing policy, their methods and results need to be
coupled with a clearer theoretical foundation and understanding of the
types of bots being measured and analyzed. Although the relevant
literature has expanded significantly in the past two years, there has
been little of the definitional debate and the theoretical work one
would expect: much of the recent theoretical and ethnographic work on
bots is not in conversation with current quantitative efforts to measure
bots and their impact. As a result, qualitative and quantitative
approaches to bot research have yet to establish a common typology for
interpreting the outputs of these research communities, thereby
requiring policymakers to undergo unwieldly synthetic work in defining
bots and their impact in their effort to pursue evidence-based policy.
As a translational effort between quantitative and qualitative research,
the typology developed in this article aims to provide a framework for
facilitating the cumulative development of shared concepts and
measurements regarding bots, media manipulation, and political
automation more generally, with the ultimate goal of providing clearer
guidance in the development of bot policy.

Beyond the conceptual ambiguities discussed is this article, there are
several other challenges that face the researchers, policymakers, and
journalists trying to understand and accurately engage with politically
relevant forms of online automation moving forward. These, most
pressingly, include imperfect bot detection methods and an overall lack
of reliable data. Future work will be required to engage deeply with the
question of what can be done to overcome these challenges of poor
measurement, data access, and---perhaps most importantly---the intricate
layers of overlapping public, corporate, and government interests that
define this issue area.

\section{References}
\label{S:6}

Abokhodair, Norah, Daisy Yoo, and David W. McDonald. 2015. ``Dissecting
a Social Botnet: Growth, Content and Influence in Twitter.'' In,
839--51. ACM.

Adams, Terrence. 2017. ``AI-Powered Social Bots.''
\emph{arXiv:1706.05143 {[}Cs{]}}, June.
{http://arxiv.org/abs/1706.05143}.

Al-Rawi, Ahmed K. 2014. ``Cyber Warriors in the Middle East: The Case of
the Syrian Electronic Army.'' \emph{Public Relations Review} 40 (3):
420--28.

Albright, Jonathan. 2017. ``Itemized Posts and Historical Engagement - 6
Now-Closed FB Pages.''

Alvisi, Lorenzo, Allen Clement, Alessandro Epasto, Silvio Lattanzi, and
Alessandro Panconesi. 2013. ``Sok: The Evolution of Sybil Defense via
Social Networks.'' In \emph{Security and Privacy (SP), 2013 IEEE
Symposium on}, 382--96. IEEE.

Bastos, M. T., and D. Mercea. 2017. ``The Brexit Botnet and
User-Generated Hyperpartisan News.'' \emph{Social Science Computer
Review}, September.

Bessi, Alessandro, and Emilio Ferrara. 2016. ``Social Bots Distort the
2016 U.S. Presidential Election Online Discussion.'' \emph{First Monday}
21 (11).

Bogost and N. Montfort, 2009. Platform Studies: Frequently Questioned
Answers. in Proceedings of the Digital Arts and Culture Conference,
Irvine CA, December 12-15.

Boshmaf, Yazan, Ildar Muslukhov, Konstantin Beznosov, and Matei Ripeanu.
2011. ``The Socialbot Network: When Bots Socialize for Fame and Money.''
In \emph{Proceedings of the 27th Annual Computer Security Applications
Conference}, 93--102. ACSAC '11. New York, NY, USA: ACM.

---------. 2013. ``Design and Analysis of a Social Botnet.''
\emph{Computer Networks}, Botnet Activity: Analysis, Detection and
Shutdown, 57 (2): 556--78.

boyd, danah. 2012. ``The Politics of Real Names.'' \emph{Communications
of the ACM} 55 (8): 29--31.

Brunton, Finn. 2012. ``Constitutive Interference: Spam and Online
Communities.'' \emph{Representations} 117 (1): 30--58.

---------. 2013. \emph{Spam: A Shadow History of the Internet}. MIT
Press.

Bu, Zhan, Zhengyou Xia, and Jiandong Wang. 2013. ``A Sock Puppet
Detection Algorithm on Virtual Spaces.'' \emph{Knowledge-Based Systems}
37 (January): 366--77.

Cao, Qiang, Michael Sirivianos, Xiaowei Yang, and Tiago Pregueiro. 2012.
``Aiding the Detection of Fake Accounts in Large Scale Social Online
Services.'' In. USENIX Association.

Cardullo, Paolo. 2015. ```Hacking Multitude' and Big Data: Some Insights
from the Turkish `Digital Coup'.'' \emph{Big Data \& Society} 2 (1):
2053951715580599.

Chen, Adrian. 2015. ``The Agency.'' \emph{The New York Times}, June.

Chu, Zi, Steven Gianvecchio, Haining Wang, and Sushil Jajodia. 2010.
``Who Is Tweeting on Twitter: Human, Bot, or Cyborg?'' In
\emph{Proceedings of the 26th Annual Computer Security Applications
Conference}, 21--30. ACM.

Coleman, E. Gabriella. 2012. ``Phreaks, Hackers, and Trolls: The
Politics of Transgression and Spectacle.'' In \emph{The Social Media
Reader}, edited by Mandiberg, Michael. New York: New York University
Press.

Davis, Clayton Allen, Onur Varol, Emilio Ferrara, Alessandro Flammini,
and Filippo Menczer. 2016. ``BotOrNot: A System to Evaluate Social
Bots.'' In \emph{Proceedings of the 25th International Conference
Companion on World Wide Web}, 273--74. 2889302: International World Wide
Web Conferences Steering Committee.

Deryugina, OV. 2010. ``Chatterbots.'' \emph{Scientific and Technical
Information Processing} 37 (2): 143--47.

Edwards, Chad, Autumn Edwards, Patric R. Spence, and Ashleigh K.
Shelton. 2014. ``Is That a Bot Running the Social Media Feed? Testing
the Differences in Perceptions of Communication Quality for a Human
Agent and a Bot Agent on Twitter.'' \emph{Computers in Human Behavior}
33: 372--76.

Ferrara, Emilio. 2017. ``Disinformation and Social Bot Operations in the
Run up to the 2017 French Presidential Election.''
\emph{arXiv:1707.00086 {[}Physics{]}}, June.
{http://arxiv.org/abs/1707.00086}.

Ferrara, Emilio, Onur Varol, C. Davis, F. Menczer, and A. Flammini.
2016. ``The Rise of Social Bots.'' \emph{Communications of the ACM} 59
(7): 96--104.

Floridi, Luciano. 2014. \emph{The Fourth Revolution: How the Infosphere
Is Reshaping Human Reality}. Oxford University Press.

Folstad, Asbjørn, and Petter Bae Brandtzaeg. 2017. ``Chatbots and the
New World of HCI.'' \emph{Interactions} 24 (4): 38--42.

Ford, Heather, Elizabeth Dubois, and Cornelius Puschmann. 2016.
``Keeping Ottawa Honest - One Tweet at a Time? Politicians, Journalists,
Wikipedians and Their Twitter Bots.'' \emph{International Journal of
Communication} 10: 4891-4914.

Franklin, Stan, and Art Graesser. 1996. ``Is It an Agent, or Just a
Program?: A Taxonomy for Autonomous Agents.'' In \emph{Intelligent
Agents III Agent Theories, Architectures, and Languages}, 21--35.
Lecture Notes in Computer Science. Springer, Berlin, Heidelberg.

Freedom House. 2016. ``Freedom on the Net Report: Ecuador.''
{https://freedomhouse.org/report/freedom-net/2016/ecuador}.

Geiger, Stuart. 2014. Bots, bespoke, code and the materiality of
software platforms. \emph{Information, Communication \& Society}
\emph{17}(3): 342--356.

Gilani, Zafar, Jon Crowcroft, Reza Farahbakhsh, and Gareth Tyson. 2017.
``The Implications of Twitterbot Generated Data Traffic on Networked
Systems.'' In \emph{Proceedings of the SIGCOMM Posters and Demos},
51--53. SIGCOMM Posters and Demos '17. New York.

Gilani, Zafar, Reza Farahbakhsh, and Jon Crowcroft. 2017. ``Do Bots
Impact Twitter Activity?'' In \emph{Proceedings of the 26th
International Conference on World Wide Web Companion}, 781--82.
International World Wide Web Conferences Steering Committee.

Gillespie, Tarleton. 2015. ``Platforms Intervene.'' \emph{Social Media +
Society} 1 (1): 2056305115580479.

Gillespie, Tarleton. 2018. \emph{Custodians of the Internet: Platforms,
Content Moderation, and the Hidden Decisions that Shape Social Media}.
New Haven: Yale University Press.

Glaser, April. 2017. ``Twitter Could Do a Lot More to Curb the Spread of
Russian Misinformation.'' \emph{Slate}, October.

Global Voices. 2017. ``Netizen Report: Vietnam Says Facebook Will
Cooperate with Censorship Requests on Offensive and `Fake' Content ·
Global Voices.''

Golumbia, David, Commercial Trolling: Social Media and the Corporate
Deformation of Democracy (July 31, 2013). SSRN.

Gorwa, Robert. 2017. ``Computational Propaganda in Poland: False
Amplifiers and the Digital Public Sphere.'' \emph{Project on
Computational Propaganda Working Paper Series: Oxford, UK}.

Gorwa, Robert, and Douglas Guilbeault. 2017. ``Tinder Nightmares: The
Promise and Peril of Political Bots.'' \emph{WIRED UK}, July.

Grimmelmann, James. 2015. ``The Law and Ethics of Experiments on Social
Media Users.'' SSRN Scholarly Paper ID 2604168. Rochester, NY: Social
Science Research Network.

Gudivada, Venkat N, Vijay V Raghavan, William I Grosky, and Rajesh
Kasanagottu. 1997. ``Information Retrieval on the World Wide Web.''
\emph{IEEE Internet Computing} 1 (5): 58--68.

Hayati, Pedram, Kevin Chai, Vidyasagar Potdar, and Alex Talevski. 2009.
``HoneySpam 2.0: Profiling Web Spambot Behaviour.'' In \emph{Principles
of Practice in Multi-Agent Systems}, 335--44.

Hogan, Bernie. 2012. ``Pseudonyms and the Rise of the Real-Name Web.''
SSRN Scholarly Paper ID 2229365. Rochester, NY: Social Science Research
Network. 

Karpf, David. 2017. ``People Are Hyperventilating over a Study of
Russian Propaganda on Facebook. Just Breathe Deeply.'' \emph{Washington
Post}.

King, G., J. Pan, and M. Roberts, 2017. How the Chinese Government
Fabricates Social Media Posts for Strategic Distraction, Not Engaged
Argument. American Political Science Review 111 (3): 484-501

Koster, Martijn. 1996. ``A Method for Web Robots Control.'' \emph{IETF
Network Working Group, Internet Draft}.

Lazarsfeld, Paul Felix, and Allen H Barton. 1957. \emph{Qualitative
Measurement in the Social Siences: Classification, Typologies, and
Indices}. Stanford University Press.

Lee, Kyumin, Brian David Eoff, and James Caverlee. 2011. ``Seven Months
with the Devils: A Long-Term Study of Content Polluters on Twitter.'' In
\emph{In AAAI Int'l Conference on Weblogs and Social Media (ICWSM)}.

Lee, Kyumin, Steve Webb, and Hancheng Ge. 2014. ``The Dark Side of
Micro-Task Marketplaces: Characterizing Fiverr and Automatically
Detecting Crowdturfing.'' In \emph{International Conference on Weblogs
and Social Media (ICWSM)}.

Leonard, Andrew. 1997. \emph{Bots: The Origin of the New Species}. Wired
Books.

Lokot, Tetyana, and Nicholas Diakopoulos. 2016. ``News Bots: Automating
News and Information Dissemination on Twitter.'' \emph{Digital
Journalism} 4 (6): 682--699.

Manjoo, Farhad, and Kevin Roose. 2017. ``How to Fix Facebook? We Asked 9
Experts.'' \emph{The New York Times}, October.

Marwick, Alice, and Rebecca Lewis. 2017. ``Media Manipulation and
Disinformation Online.'' \emph{Data and Society Research Institute
Report}.

Maus, Gregory. 2017. ``A Typology of Socialbots (Abbrev.).'' In
\emph{Proceedings of the 2017 ACM on Web Science Conference}, 399--400.
WebSci '17. New York, NY, USA: ACM.

Metaxas, Panagiotis T, and Eni Mustafaraj. 2012. ``Science and Society.
Social Media and the Elections.'' \emph{Science} 338 (6106): 472--73.

Mitter, Silvia, Claudia Wagner, and Markus Strohmaier. 2014.
``Understanding the Impact of Socialbot Attacks in Online Social
Networks.'' \emph{arXiv Preprint arXiv:1402.6289}.

Moore, Tyler, and Ross Anderson. 2012. ``Internet Security.'' In
\emph{The Oxford Handbook of the Digital Economy}. Oxford University
Press.

Munger, Kevin. 2017. ``Tweetment Effects on the Tweeted: Experimentally
Reducing Racist Harassment.'' \emph{Political Behavior} 39 (3): 629--49.

Napoli, Philip, and Robyn Caplan. 2017. ``Why Media Companies Insist
They're Not Media Companies, Why They're Wrong, and Why It Matters.''
\emph{First Monday} 22 (5).

Neff, Gina, and Dawn Nafus. 2016. \emph{Self-Tracking}. MIT Press.

Neff, G., and P. Nagy, 2016. Talking to Bots: Symbiotic Agency and the
Case of Tay. International Journal of Communication 10: 4915-4931

Olston, Christopher, and Marc Najork. 2010. ``Web Crawling.''
\emph{Foundations and Trends in Information Retrieval} 4 (3): 175--246.

Pant, Gautam, Padmini Srinivasan, and Filippo Menczer. 2004. ``Crawling
the Web.'' In \emph{Web Dynamics: Adapting to Change in Content, Size,
Topology and Use}, edited by Mark Levene and Alexandra Poulovassilis.
Springer Science \& Business Media.

Pearce, Katy E. 2015. ``Democratizing Kompromat: The Affordances of
Social Media for State-Sponsored Harassment.'' \emph{Information,
Communication \& Society} 18 (10): 1158--74.

Phillips, Whitney. 2015. \emph{This Is Why We Can't Have Nice Things:
Mapping the Relationship Between Online Trolling and Mainstream
Culture}. Cambridge, Massachusetts: MIT Press.

Ratkiewicz, Jacob, Michael Conover, Mark Meiss, Bruno Gonçalves, Snehal
Patil, Alessandro Flammini, and Filippo Menczer. 2011. ``Truthy: Mapping
the Spread of Astroturf in Microblog Streams.'' In \emph{Proceedings of
the 20th International Conference Companion on World Wide Web}, 249--52.

Read, Max. 2017. ``Does Even Mark Zuckerberg Know What Facebook Is?''
\emph{New York Magazine}.

Rodríguez-Gómez, Rafael A, Gabriel Maciá-Fernández, and Pedro
García-Teodoro. 2013. ``Survey and Taxonomy of Botnet Research Through
Life-Cycle.'' \emph{ACM Computing Surveys (CSUR)} 45 (4): 45.

Rosenberg, Yair. 2017. ``Opinion Confessions of a Digital Nazi Hunter.''
\emph{The New York Times}, December.

Sansonnet, Jean-Paul, David Leray, and Jean-Claude Martin. 2006.
``Architecture of a Framework for Generic Assisting Conversational
Agents.'' In \emph{Intelligent Virtual Agents}, 145--56. Lecture Notes
in Computer Science. Springer, Berlin, Heidelberg.

Sartori, Giovanni. 1970. ``Concept Misformation in Comparative
Politics.'' \emph{American Political Science Review} 64 (4): 1033--53.

Savage, Saiph, Andres Monroy-Hernandez, and Tobias Hollerer. 2015.
``Botivist: Calling Volunteers to Action Using Online Bots.''
\emph{arXiv Preprint arXiv:1509.06026}.

Shane, Scott. 2017. ``The Fake Americans Russia Created to Influence the
Election.'' \emph{The New York Times}.

Shao, Chengcheng, Giovanni Luca Ciampaglia, Onur Varol, Alessandro
Flammini, and Filippo Menczer. 2017. ``The Spread of Fake News by Social
Bots.'' \emph{arXiv:1707.07592 {[}Physics{]}}, July.

Stieglitz, Stefan, Florian Brachten, Björn Ross, and Anna-Katharina
Jung. 2017. ``Do Social Bots Dream of Electric Sheep? A Categorisation
of Social Media Bot Accounts.'' \emph{arXiv:1710.04044 {[}Cs{]}},
October. 

Stretch, Colin. 2017. ``Facebook to Provide Congress with Ads Linked to
Internet Research Agency.'' \emph{FB Newsroom}.

Suárez-Serrato, Pablo, Margaret E. Roberts, Clayton Davis, and Filippo
Menczer. 2016. ``On the Influence of Social Bots in Online Protests.''
In \emph{Social Informatics}, 269--78. Lecture Notes in Computer
Science. Springer.

Subrahmanian, V. S., Amos Azaria, Skylar Durst, Vadim Kagan, Aram
Galstyan, Kristina Lerman, Linhong Zhu, et al. 2016. ``The DARPA Twitter
Bot Challenge.'' \emph{Computer} 49 (6): 38--46.

Thomas, Kurt, Chris Grier, and Vern Paxson. 2012. ``Adapting Social Spam
Infrastructure for Political Censorship.'' In \emph{LEET}.

Tsvetkova, Milena, Ruth García-Gavilanes, Luciano Floridi, and Taha
Yasseri. 2017. ``Even Good Bots Fight: The Case of Wikipedia.''
\emph{PLOS ONE} 12 (2): e0171774.

Tucker, Joshua A, Yannis Theocharis, Margaret E Roberts, and Pablo
Barberá. 2017. ``From Liberation to Turmoil: Social Media and
Democracy.'' \emph{Journal of Democracy} 28 (4): 46--59.

Twitter Policy. 2017. ``Update: Russian Interference in 2016 US
Election, Bots, \& Misinformation.''

---------. 2018. ``Update on Twitter's Review of the 2016 U.S.
Election.''

Tworek, Heidi. 2017. ``How Germany Is Tackling Hate Speech.''
\emph{Foreign Affairs}.

Verkamp, John-Paul, and Minaxi Gupta. 2013. ``Five Incidents, One Theme:
Twitter Spam as a Weapon to Drown Voices of Protest.'' In \emph{FOCI}.

Vosoughi, Soroush, Deb Roy, and Sinan Aral. 2018. The spread of true and
false news online. \emph{Science} 359 (6380): 1146--1151.

Wang, Selina. 2018. ``California Would Require Twitter, Facebook to
Disclose Bots.'' \emph{Bloomberg}, April.

Weedon, Jen, William Nuland, and Alex Stamos. 2017. ``Information
Operations and Facebook.'' \emph{Facebook Security White Paper.}.

Weinstein, Lauren. 2003. ``Spam Wars.'' \emph{Communications of the ACM}
46 (8): 136.

Weizenbaum, Joseph. 1966. ``ELIZA---a Computer Program for the Study of
Natural Language Communication Between Man and Machine.''
\emph{Communications of the ACM} 9 (1): 36--45.

Woolley, Samuel C. 2016. ``Automating Power: Social Bot Interference in
Global Politics.'' \emph{First Monday} 21 (4).

Woolley, Samuel C., and Philip N. Howard. 2016. ``Political
Communication, Computational Propaganda, and Autonomous Agents ---
Introduction.'' \emph{International Journal of Communication} 10
(October): 4882--90.

Wojcik, Stefan, Solomon Messing, Aaron Smith, Lee Rainie, and Paul
Hitlin. 2018. ``Bots in the Twittersphere.'' Pew Research Center.

Yang, Zhi, Christo Wilson, Xiao Wang, Tingting Gao, Ben Y Zhao, and
Yafei Dai. 2014. ``Uncovering Social Network Sybils in the Wild.''
\emph{ACM Transactions on Knowledge Discovery from Data (TKDD)} 8 (1):
1--29.

Yao, Yuanshun, Bimal Viswanath, Jenna Cryan, Haitao Zheng, and Ben Y.
Zhao. 2017. ``Automated Crowdturfing Attacks and Defenses in Online
Review Systems.'' \emph{arXiv:1708.08151 {[}Cs{]}}.

\end{document}